# Creativity in the Age of AI: Evaluating the Impact of Generative AI on Design Outputs and Designers' Creative Thinking

YUE FU, Information School, University of Washington, US
HAN BIN, Information School, University of Washington, US
TONY ZHOU, Paul G. Allen School of Computer Science & Engineering, University of Washington, US
MARX WANG, Information School, University of Washington, US
YIXIN CHEN, Information School, University of Washington, US
ZELIA GOMES DA COSTA LAI, Human-Centered Design and Engineering, University of Washington, US
JACOB O. WOBBROCK, Information School, University of Washington, US
ALEXIS HINIKER, Information School, University of Washington, US

As generative AI (GenAI) increasingly permeates design workflows, its impact on design outcomes and designers' creative capabilities warrants investigation. We conducted a within-subjects experiment where we asked participants to design advertisements both with and without GenAI support. Our results show that expert evaluators rated GenAI-supported designs as more creative and unconventional ("weird") despite no significant differences in visual appeal, brand alignment, or usefulness, which highlights the decoupling of novelty from usefulness—traditional dual components of creativity—in the context of GenAI usage. Moreover, while GenAI does not significantly enhance designers' overall creative thinking abilities, users were affected differently based on native language and prior AI exposure. Native English speakers experienced reduced relaxation when using AI, whereas designers new to GenAI exhibited gains in divergent thinking such as idea fluency and flexibility. These findings underscore the variable impact of GenAI on different user groups, suggesting the potential for customized AI tools to better meet diverse user needs.

## 1 Introduction

Creativity, commonly defined as the ability to generate novel and useful ideas [97], is increasingly recognized as a pivotal skill in modern society. Research has highlighted creativity's significant role in improving problem-solving, fostering innovation, boosting productivity, and contributing to personal socio-emotional well-being, satisfaction, and life success [7, 77, 92, 95, 96], highlighting its significance for both individual development and societal advancement.

The design industry is a key environment for applying creativity, where the constant generation of fresh and innovative ideas is essential [27]. Creativity drives the production of novel and useful outputs [12] that meet client needs and enable businesses to stand out in competitive markets. In 2021, the global design industry was estimated at $162 billion and consistently growing [44].

To address the increasing demand for innovative design solutions, creativity support tools (CSTs) have become widely used in the design industry. For example, over 90% creative professionals worldwide use Adobe Photoshop,

Authors' Contact Information: Yue Fu, chrisfu@uw.edu, Information School, University of Washington, Seattle, Washington, US; Han Bin, bh193@uw.edu, Information School, University of Washington, Seattle, Washington, US; Tony Zhou, tyzhou05@uw.edu, Paul G. Allen School of Computer Science & Engineering, University of Washington, Seattle, Washington, US; Marx Wang, marxwang@uw.edu, Information School, University of Washington, Seattle, Washington, US; Yixin Chen, Information School, University of Washington, Seattle, Washington, US, yixin7@uw.edu; Zelia Gomes da Costa Lai, Human-Centered Design and Engineering, University of Washington, Seattle, Washington, US, zelia@uw.edu; Jacob O. Wobbrock, Information School, University of Washington, Seattle, Washington, US, wobbrock@uw.edu; Alexis Hiniker, Information School, University of Washington, Seattle, Washington, US, .

which boasts an estimated 23 million monthly users [60]. Canva, an online graphic editing tool launched a decade ago, has attracted 170 million monthly users worldwide in early 2024 [22]. These CSTs are designed to enhance both the creative person and the creative activity [45], aiming to "make people more creative more often" [89] and facilitating creative activities by improving the processes, usability, and workflows involved in creative work [81].

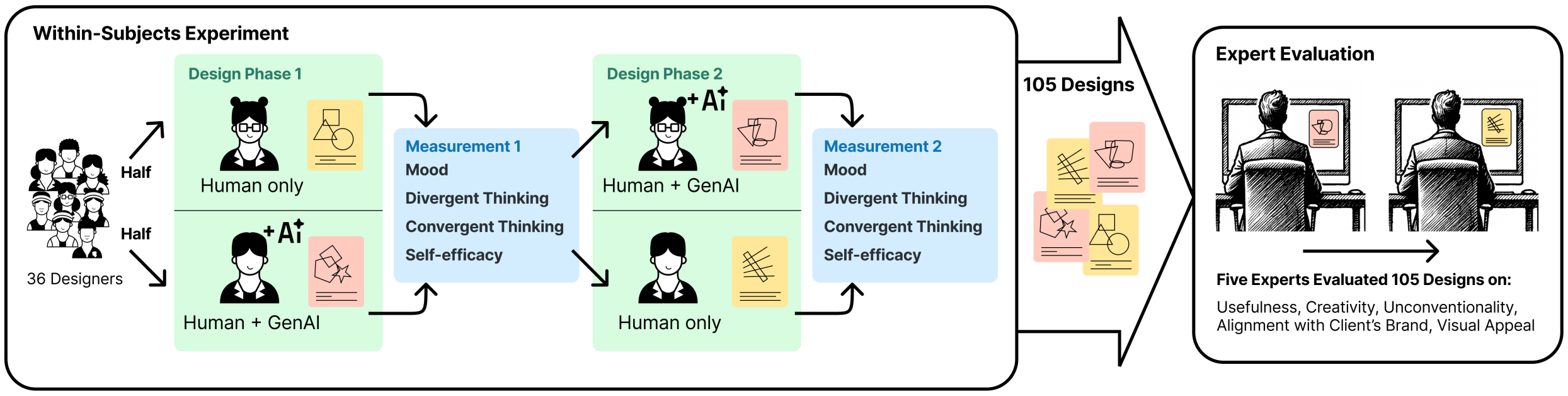

Fig. 1. Our within-subjects experiment design and expert evaluation procedure. We recruited 36 designers to create social media advertisements promoting a client's organization. Participants were randomly assigned to two groups and completed two design sessions in a counterbalanced order. In the first session, one group designed independently (human-only), while the other collaborated with a GenAI tool (GenAI-supported). After the first session, researchers measured participants' mood, self-efficacy, divergent thinking with the Alternative Uses Test [4], and convergent thinking with the Anagram Test [14]. In the second session, the groups switched conditions: the human-only group used the GenAI tool, and the GenAI-supported group designed independently. The same measurements were conducted after the second session. Across both sessions, a total of 105 eligible designs were collected. Five experts, blind to the experimental conditions, evaluated the designs based on their usefulness, creativity, unconventionality ("weirdness"), alignment with the client's brand, and visual appeal.

Recently, GenAI-powered CSTs have gained significant traction for their potential to transform how designers work, offering new opportunities for creative expression and efficiency. These GenAI tools claim to accelerate the creative process by automating repetitive tasks [8], suggesting innovative design concepts [99], and providing real-time feedback [105]. Text-to-image models like OpenAI's DALL-E [69] and Midjourney [65] can generate high-quality visual content from text descriptions, enabling designers to quickly prototype ideas [31]. For example, Adobe's Creative Cloud suite has incorporated GenAI features like "*Generative Fill*" [1] and "*Neural Filters,*" marketed for their ease of use: "*In some cases, with just one click. In other cases, it gets you 80% of the way there*" [2].

However, as the design industry increasingly relies on these tools, questions emerge:

- **RQ1:** Are these generative AI creative support tools truly delivering on their promises of enhancing designers' creativity?
- **RQ2:** If so, how do they affect design output?
- **RQ3:** What is their influence on the experience of conducting design work, specifically, their effects on a designer's mood, inherent creativity, and self-efficacy?

To investigate these questions, we conducted a lab experiment with 36 designers from diverse design backgrounds. Participants were asked to complete a common real-world design task: creating advertising material for a social media campaign to promote a client organization. The experiment consisted of two sessions. In one session, participants designed the advertisement with the support of GenAI (GenAI-supported condition, text-to-image model DALL-E 2 using the ChatGPT web app), while in the other, they designed without GenAI support (human-only condition). All participants completed both sessions, with the order of conditions counterbalanced to control for any order effects. After each session, we measured their mood (assessed through self-reported energy, relaxation, and pleasure), inherent creative ability (with one measure for divergent thinking and another for convergent thinking, as detailed in Section 3.2), and self-reported self-efficacy. Each participant submitted one or two designs per session. In total, we collected 105 eligible designs, which were evaluated by five expert reviewers from the client organization who were blind to conditions. The experts rated each design for its visual appeal, alignment with the client's brand, creativity, unconventionality (operationalized as "weirdness"), and perceived usefulness for a social media campaign. Additionally, experts provided optional qualitative feedback on each design, noting aspects they liked or disliked and offering general comments.

Our findings show that experts rated GenAI-supported designs as more creative and more unconventional than human-only designs. However, there were no significant differences between the two conditions in perceived usefulness, visual appeal, or brand representation. Qualitative analysis of expert feedback indicated that while GenAI-supported designs were considered more creative and colorful, they were also more likely to be criticized for being "too busy," causing visual overload, or having unnatural or abnormal composition. Experts also noted that, although GenAI-supported designs effectively represented technology and human diversity, some failed to align with social norms and could convey inappropriate messages.

When understanding designers' experiences, we found that using GenAI for a single 25-minute design session had no significant overall impact on designers' inherent creative abilities, mood, or self-efficacy. However, further subgroup analysis revealed that GenAI's effects varied based on designers' prior exposure to GenAI tools and native language. Specifically, for participants with no prior GenAI design tool exposure, using GenAI significantly increased divergent thinking as measured by idea fluency and flexibility. However, there was no similar effect on participants with prior GenAI design tool exposure, suggesting that exposure to novel design ideas from GenAI for new users is likely to increase their divergent thinking abilities. Additionally, native English speakers experienced a significant decrease in relaxation levels with GenAI use, while non-native speakers did not, possibly due to individualist and collectivist cultural factor influences.

Our study provides empirical evidence of GenAI's influence on design outputs and designers' experiences. We discuss the observed decoupling effect GenAI has on traditional two components of creativity: novelty and usefulness. In addition, we argue for the intrinsic value of creativity, underscoring the importance of designing GenAI tools that enhance human creativity rather than diminish it through automation. Lastly we discuss the implications for developing the next generation of GenAI creativity support tools tailored to diverse user groups.

## 2 Related Work

### 2.1 Creativity: Aspects, Measurement, and Enhancement

The exploration of human creativity has intrigued scholars since ancient times [50]. One famous framework, the 4P theory of creativity, delineates creativity into four aspects: *Person* (individual traits like personality, cognitive abilities, and motivation), *Process* (stages involved in creative thinking such as ideation and evaluation), *Product* (the outcome of creativity marked by originality and impact), and *Press* (environmental factors influencing creativity) [83]. Our research primarily focuses on evaluating GenAI's impact on the *Person* and *Product* aspects of this framework.

| Constructs: | Human Inherent Creativity - Measurement 1, 2 | Design Product (Output) Creativity - Expert Evaluation |
|---|---|---|
| **Measurements:** | • Divergent thinking - Alternative Use Test<br>Coding to three facets: fluency, flexibility, originality<br>• Convergent Thinking - Anagram Test | • Usefulness ("how well will this ad perform?")<br>• Creativity ("This design is creative")<br>• Unconventionality ("This design is weird") |

Fig. 2. We operationalized creativity into two components: human inherent creativity and design product (output) creativity. Human inherent creativity was measured twice for each participant: once after designing with Generative AI (GenAI) and once after designing independently (human-only). This was assessed using the Alternative Uses Test [4] (for divergent thinking, coded into three facets: fluency, flexibility, and originality) and the Anagram Test [14] (for convergent thinking). The second component, design product creativity, focused on evaluating the creative outputs generated during the design sessions. Five experts evaluated the designs' usefulness ("how well will this ad perform?", 11-point scale), perceived creativity ("this design is creative", strongly agree to strongly disagree 7-point Likert scale), and unconventionality ("this design is weird", strongly agree to strongly disagree 7-point Likert scale).

Creativity from the *Person* perspective involves key components such as divergent and convergent thinking [103]. Divergent thinking is defined as the measure of the varied potential of creative thought and focuses on the processes of creative ideation [85, 87]. Divergent thinking relates to the ability to generate multiple creative ideas, typically assessed via tests like the Alternative Uses Test (AUT), which measures how creatively participants can think of non-common uses for common objects like a "paper clip" [40, 56]. The AUT evaluates a person's creativity across three facets: fluency, the quantity of ideas generated; flexibility, the variety of idea categories; originality, uniqueness, or novel the ideas are compared to the ideas generated by other participants. On the other hand, convergent thinking involves narrowing down these possibilities to find a single correct solution [13–15]. One common test of convergent thinking is the Anagram Test, where participants rearrange letters to form words, for example, "arc" translates to "car". With respect to the *Product* aspect, creativity is evaluated by the novelty and usefulness of the outputs, often externally by reviewers rather than self-assessed by creators [26].

A number of approaches have been explored to enhance creativity, such as psychological interventions, educational programs, environmental designs, and technological tools [42, 88]. In addition, research has found that psychological factors, such as mood [11] and self-efficacy [78, 94], influence creativity. For example, studies have found that stimulating mental and physical relaxation through progressive muscle relaxation, yoga, stretching, and music relieves anxiety and leads to increases in both convergent and divergent creative ability [37, 52]. Continuing physiological research has explored the role of physical movement activity in enhancing creativity and found that activities such as walking [73], stair-climbing [51], and aerobic exercise [104] can positively impact creative thinking.

In our current study, we examine how GenAI impacts creativity, focusing its influence on both designers' divergent and convergent thinking abilities and evaluating the creative quality of design products.

## 2.2 Creativity Support Tools

With the advancement of computer technology, modern software tools have been created to support creative processes, which have been classified broadly as *creativity support tools* (CSTs) [33, 45, 89, 90]. CSTs are designed to support both creativity in individuals and collaboration in teams, aiming to *"empower users to not only be more productive, but more* *innovative"* [82]. Today, CSTs are widely used in digital design, with tools like Adobe's Creative Suite (Photoshop, Illustrator, etc.) [60], Canva [22] as the most popular.

Recently, more and more CSTs, especially these popular ones, are rapidly incorporating generative AI features to assist designers [1, 2, 21, 32]. They promise to provide support across design stages, from idea exploration [47] to detailed prototyping [53].

Although CSTs are promising in increasing usability and workflows in designers' creative work, one review study has shown that CSTs have notably 1) lacked expert evaluations of both the usefulness and usability of the tools, 2) included little theoretical grounding, and 3) focused on usability rather than creativity, which the authors argue is crucial for future work to consider [81]. Regarding new GenAI-powered CSTs, there remains a scarcity of prior work validating their effectiveness and impact on design outputs. Our study explores the influence of these GenAI-enhanced CSTs on designers' work and their creative outputs, aiming to fill the gap in current research.

### 2.3 Generative AI and Creativity

Recent advancements in AI, particularly through multimodal large language models [3, 70], have significantly broadened the scope for AI-assisted creativity. GenAI can produce novel and diverse outputs in various domains, including writing, communication, art, music, and design [5, 35, 57, 64, 106]. Some research claims these models not only generate content but also enhance human creativity by providing diverse stimuli and suggesting unconventional combinations, which can disrupt conventional thinking patterns [38, 98]. Others caution that an overreliance on AI might constrain human creativity by promoting conformity or limiting exploration [46, 48].

One promising approach to human-AI collaboration that researchers have noted is the concept of mixed-initiative co-creativity [26, 63, 101]. In this paradigm, human and AI agents work together, each contributing according to their strengths. Studies have investigated how mixed-initiative systems can support various aspects of the creative process, such as idea generation, conceptual exploration, and design refinement [25, 55, 59].

However, the impact of generative AI on mixed-initiative creativity is still a highly contested topic of ongoing research. One study found that access to GenAI worsened creative performance for participants whose creative potential was above the sample median [107]. While some studies suggest that generative AI can enhance individual creativity by providing inspiration and facilitating exploration [41, 67], others have found that it may reduce conceptual diversity at a collective level [28]. In a recent large-scale experiment study, researchers found that after participants were exposed to AI-generated design ideas, collective creative diversity was improved, but not individual creativity: ideas were different and broader, but not necessarily better [10].

Moreover, most existing studies on AI's impact on creativity have focused on textual outputs using text-to-text models, like writing essays [75], fictional stories [28], or imaginative and experimental scenarios [9, 10]. There is a notable gap in research applying AI to real-world text-to-image design tasks and evaluating these outputs to assess their creativity.

Our study aims to bridge this gap by focusing on a real-world design task supported by text-to-image GenAI models, providing a perspective on the role of GenAI in human creativity in practical, applied settings. This exploration is crucial for understanding the effect of GenAI on designers' inherent creativity and the creativity of the perceived design products.

## 3 Method

We conducted a within-subjects experimental study, in which 36 participants designed social media advertisements with and without support from GenAI. All participants participated in both sessions, with session order counterbalanced. We assessed participants' creativity after each design session, and reviewers blind to conditions evaluated participants' design outputs for their creativity, usefulness, and appeal, among other measures. Here, we describe these procedures in detailsee Fig 1.

### 3.1 Participants

We recruited 36 participants via professional and academic communication channels (Slack, Discord, etc.), email lists, and fliers posted on a large public university campus in the United States. Through an initial screening survey, we asked participants to report their design experience, creative capability, native language, prior experience using AI for design work, and demographic data. Participants were over-representative of young people, with 94% ($N = 34$) under the age of 30, and of women (81%, $N = 29$). They had a diverse range of AI design tool experience with 67% ($N = 24$) having at least some experience using GenAI for design tasks. Each participant received a $40 Amazon gift card for completing the study.

### 3.2 Materials and Apparatus

All design tasks were conducted via an in-lab-controlled environment. Participants in GenAI-supported group were asked to use the ChatGPT web app [71], powered by the DALL-E2 text-to-image model [80] on their computer, to generate and modify designs (we provided participants with ChatGPT Plus accounts). For the human-only group, they were asked to design individually using the tools they like. Participants' assessments and post-procedure expert evaluation were conducted using Qualtrics, an online survey tool [79].

We performed three types of assessment: evaluations of participants' intrinsic creativity, self-report measures of participants' mental state (mood, self-efficacy), and third-party evaluations of participants' design outputs. Here, we describe the materials used to conduct each of these evaluations.

- ***Participant Creativity.*** To measure participants' inherent creativity, we used the Alternative Uses Test (AUT) [4], an established measure of divergent thinking, and the Anagram Test [14], an established measure of convergent thinking. In the AUT, a participant brainstorms unconventional uses for a commonplace item, such as a bottle or table, with a two-minute time limit. This is then repeated with a second item. The total number of uses the participant lists across both objects is their final score. The specific instructions used were as follows:

  > "*You will be given two everyday objects. Your task is to think of as many different uses for this object as possible beyond its common or intended use. Try to come up with original, unusual, diverse, and creative uses in your responses as you can. For example, if the object given to you is a paperclip, you might think of using it to: 1) eject the SIM card from a phone, 2) be material for a tiny sculpture, 3) clean small crevices. Remember, the more creative and unusual, the better!*"

  To assess convergent thinking, we used the Anagram Test [14]. In this task, the participant rearranges the letters of a prompt word (e.g., "lamp") into a different word (e.g., "palm"). They have three minutes to rearrange as many prompt words as possible into their anagrams; the number of completed words is their final score.

- ***Participant Mental State.*** In keeping with prior work [51], we measured participant mood via a short, self-report survey with three subscales: relaxation, pleasure, and energy level. We also measured participant self-efficacy through a self-report measure developed by Dow et al. [30].
- ***Quality of Design Output Evaluated by Experts.*** To evaluate the quality of a design by experts, we created a five-question survey to be completed by five expert reviewers blind to experiment condition. This included an 11-point scale question about the usefulness of the design ("*In a social media ad campaign, how well will this ad perform?*"), and four 7-point "strongly degree" to "strong agree questions" to measure designs' creativity ("*This design is creative*"), weirdness ("*This design is weird*'), representing of the client ("*This design effectively represents [anonymous organization for which it was designed],*") and visual appeal ("*This design is visually appealing*"). In addition, the survey included three free-response textboxes where the reviewer could optionally note their likes, dislikes, and explanations for their ratings.

### 3.3 Procedures

The study consisted of a two-phase, within-subjects experiment, which lasted approximately 90 minutes. At the start of the session, the participant was informed that they would design several advertisements to be used in actual advertising campaigns on social media platforms like Facebook and Instagram. To encourage participants to do their best work, we informed them that the creators of the top three designs would receive an additional $40 gift card bonus. Participants then completed each of two study phases:

- *Phase 1:* A randomly selected half of all participants were first assigned to design individually, with the second half assigned to design with GenAI. In each case, participants were tasked with creating a $1080 \times 1080$ pixel advertisement for a social media advertising campaign to promote a newly launched research group website. Examples of social media advertisements on various devices were shown for reference. The GenAI group was instructed to use ChatGPT for design support but was also permitted to use other tools to correct inaccuracies in text generated by AI or to clip images as needed. The human-only group worked without AI assistance. In each case, participants had 25 minutes to produce one to two designs. After completing their design task, participants reported their mood and sense of self-efficacy; this process typically took less than 30 seconds. The researcher then assessed the participant's divergent thinking via the AUT, followed by convergent thinking via the Anagram Test. All assessments were done using Qualtrics. After these assessments, participants took a break for 1-3 minutes before moving on to Phase 2.
- *Phase 2.* This phase mirrored the first, with participants completing the condition they did not complete initially: those who used ChatGPT initially (GenAI-supported condition) now designed individually (human-only condition) and vice versa. After completing the second design session, participants were once again evaluated on their mood, self-efficacy, divergent thinking via the AUT, and their convergent thinking via the Anagram Test. The everyday objects and the prompt words used during this second measurement phase differed from those used during the first measurement phase. AUT objects were randomly assigned, such that each participant saw each exactly once and the condition in which they encountered it was random.

Participants' designs formed a dataset of design outputs which were then evaluated by blind reviewers. To conduct this evaluation, we recruited five members of the research group that the advertising campaign aimed to promote, who were therefore reasonable "clients" for this design work. Each reviewer used the survey we developed (see Section 3.2) to evaluate the entire set of designs produced by participants with no knowledge of who created the design or the condition (i.e., GenAI-supported or human-only) in which it was created. Evaluating all designs took approximately 45-75 minutes, with breaks allowed as needed. Reviewers received a $40 gift card as a thank-you for their time.

### 3.4 Design and Analysis

Our study had one within-subjects factor, design support, with two levels: GenAI-supported and human-only. We also examined past experience with GenAI (with prior GenAI design tool exposure, without prior GenAI design tool exposure) and English language proficiency (native language, non-native language).

The study's protocol was preregistered. We initially planned to analyze the influence of GenAI on participants with varying self-reported design skills. However, we abandoned this plan because: 1) our recruiting did not yield a participant pool with sufficiently diverse design expertise; 2) the survey question used to assess self-reported design skills lacked a clear benchmark for comparison; 3) a post-study independent samples Wilcoxon signed-rank test revealed that self-reported design skills did not significantly affect the perceived usefulness of designs ($Z = -0.238, p = 0.812$), or in other words, high self-reported designers do not produce better design outcomes. Consequently, we shifted our approach to examine GenAI's impact across the entire participant group and within subgroups that are more accurately defined, such as native language and prior GenAI design tool exposure.

One researcher coded all the AUT test scores for three facets: fluency, flexibility, and originality using the counting method in [4]. A Wilcoxon signed-rank test was used to test for changes in mood (relaxation, pleasure, energy), and self-efficacy. The same test was used to evaluate the differences between conditions in design outcomes such as usefulness, creativity, unconventionality, visual appeal, and alignment with the client's brand. To conduct our qualitative analysis of open-ended comments in experts' evaluations, one researcher conducted a thematic analysis [17] of all comments provided by the five design experts. The researcher initially added detailed notes alongside each comment to capture observations. Following this initial open coding, the researcher refined the codes through an iterative process, grouping similar codes and collapsing them into broader, more coherent themes and replying revised codes to the full dataset.

## 4 Results

In total, participants generated 110 designs (54 designs in the human-only condition and 56 designs in GenAI-supported condition). Five designs were filtered out for non-compliance with the study requirements. This resulted in a final count of 105 eligible designs—52 human-only and 53 GenAI-supported—for design evaluation. The average number of designs per participant was 1.44 in the human-only condition and 1.47 in the GenAI-supported condition.

Our primary, preregistered analysis found that experts evaluated GenAI-supported designs as more creative and unconventional ("weird") than human-only designs, but designs did not significantly differ between conditions in their perceived usefulness, visual appeal, or alignment with the client's organization. With respect to designers' creativity and mental state, our primary analyses showed no significant impact of GenAI on participants' inherent creative abilities, mood, or self-efficacy. However, our further exploratory subgroup analysis revealed differences in condition depending on designers' prior exposure to GenAI and native language. We present our quantitative results below.

### 4.1 Usefulness, Creativity, Unconventionality of Design Products

GenAI-supported designs were perceived by experts as significantly more creative and unconventional than human-only designs, though we observed no difference between conditions in usefulness ($Z = -0.238$, $p = .812$), visual appeal ($Z = -0.456$, $p = .649$), or alignment with the client's organization ($Z = -1.213$, $p = .225$). Average creativity ratings for GenAI-supported designs ($M = 3.909$, $SD = 1.389$) were rated 13% more creative than the designs from the human-only condition ($M = 3.457$, $SD = 1.359$, $Z = -3.318$, $p < .001$). This difference was even more pronounced with respect to designs' perceived unconventionality. Average unconventionality (operationalized as "weirdness") of GenAI-supported designs ($M = 4.262$, $SD = 1.328$) was 23.3% more unconventional than the designs from human-only condition ($M = 3.457$, $SD = 1.250$, $Z = -5.958$, $p < .001$).

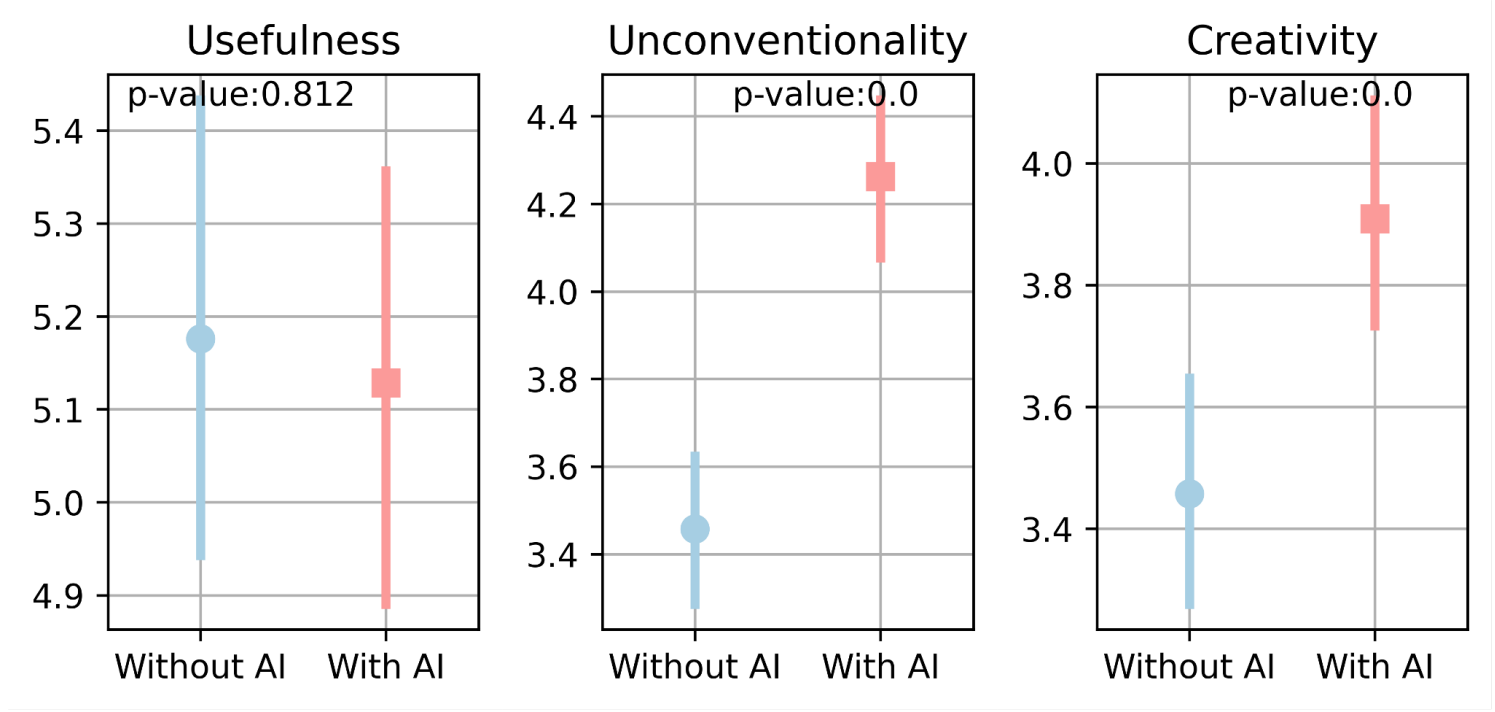

Fig. 3. Analysis of design outputs in GenAI-supported vs. human-only conditions. There is no significant difference in the perceived usefulness of designs between the conditions. However, designs supported by GenAI were rated significantly more creative ($p < 0.001$) and unconventional ($p < 0.0001$). Error bars represent 95% confidence intervals.

### 4.2 Participants Mood, Self-efficacy, and Inherent Creativity

We found that there were no significant differences between conditions in mood (relaxation, pleasure, energy), self-efficacy, inherent divergent thinking (AUT), or convergent thinking (Anagram). However, further analysis revealed significant differences between conditions within select subgroups, namely, those without prior GenAI design experience and those who were native English speakers.

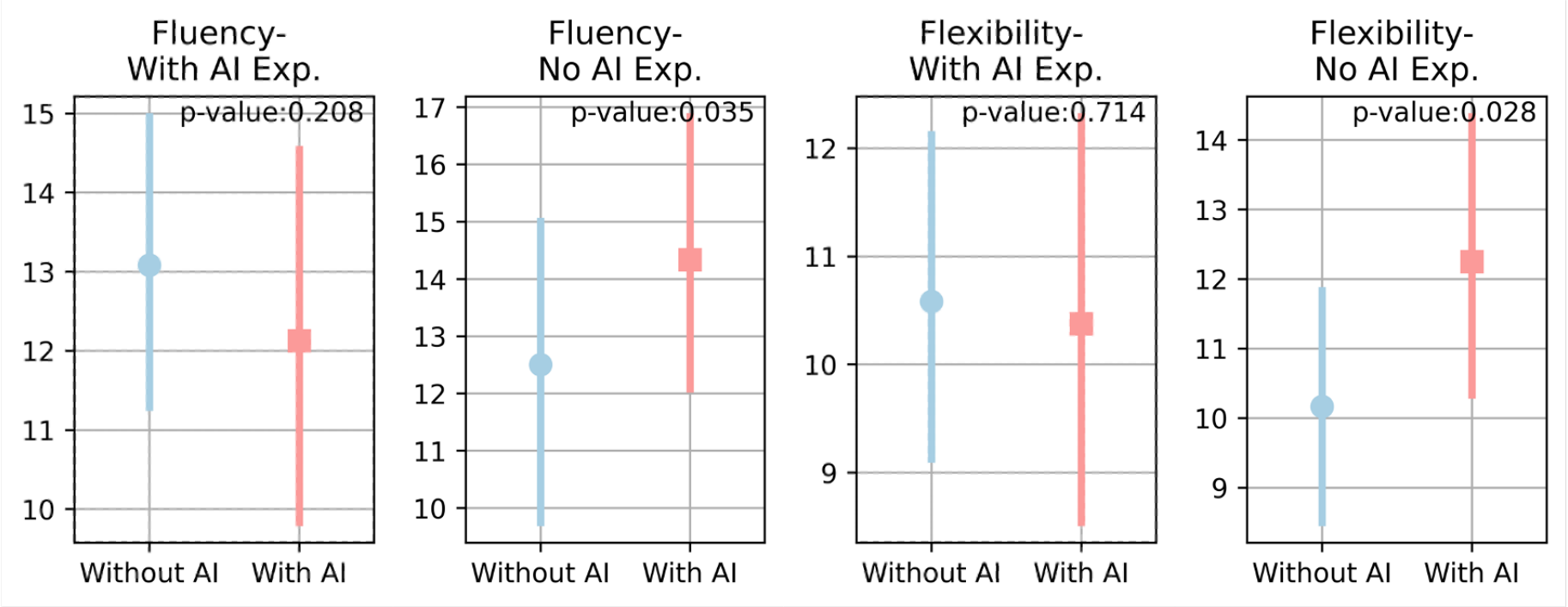

Fig. 4. Analysis of the impact of GenAI on divergent thinking measured by idea fluency and flexibility for subgroups with and without prior GenAI experience. For participants with no prior GenAI design tool experience, GenAI significantly increased idea fluency ($p < .05$) and flexibility ($p < .05$) compared to the human-only condition. No significant improvement in fluency or flexibility was observed for participants with prior GenAI experience. Error bars represent 95% confidence intervals.

*4.2.1 Subgroup: Past Experience with GenAI (with prior GenAI design tool experience, without prior GenAI design tool experience).* For participants without prior GenAI design tool experience, using GenAI for a 25-minute design session increased their divergent thinking ability. Our analysis showed a significant increase in idea fluency score ($M = 14.333$, $SD = 4.384$) and flexibility score ($M = 12.250$, $SD = 3.562$) after the GenAI-supported design session, as compared to the human-only design session (fluency, $M = 12.500$, $SD = 4.573$, $Z = -2.112$, $p < .05$; flexibility, $M = 10.167$, $SD = 3.131$, $Z = -2.203$, $p < .05$). However, for participants with prior GenAI design tool experience, there was no significant difference in divergent thinking between conditions. The experiment condition did not affect other measurements, such as convergent thinking, self-efficacy, or mood for either non-native and native speakers.

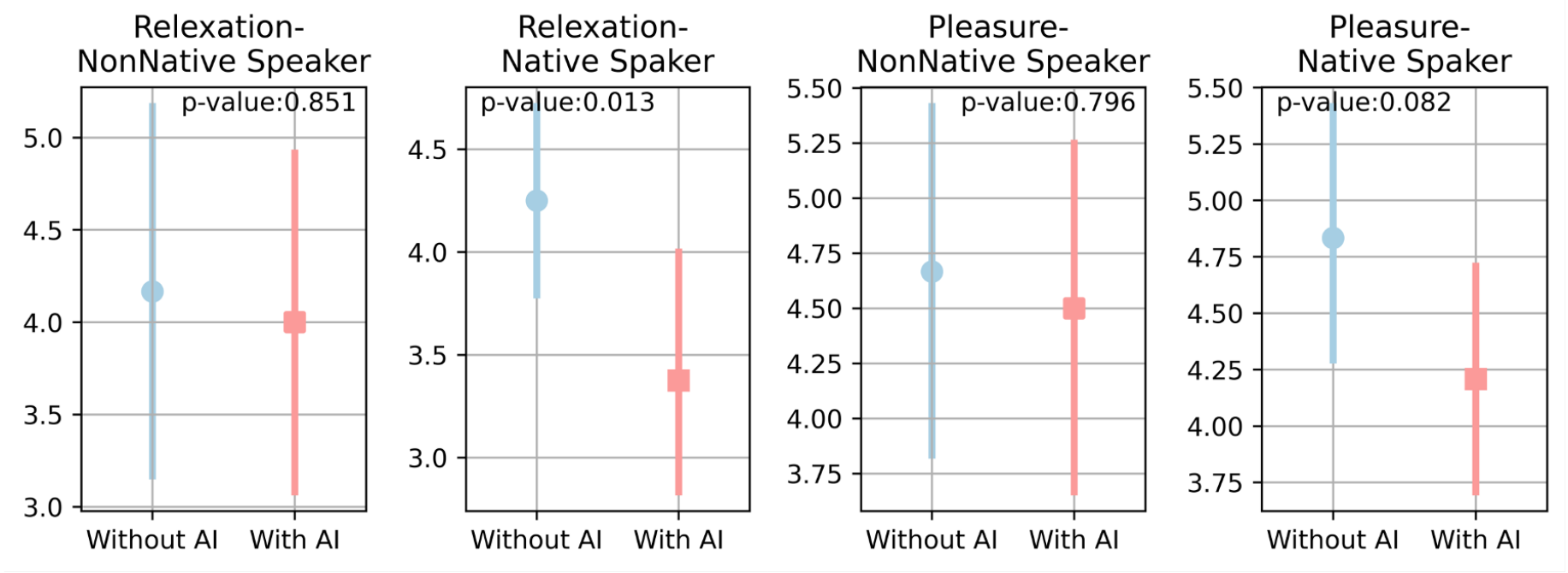

Fig. 5. Analysis of the impact of GenAI on relaxation and pleasure levels for native and non-native English speakers. Native English speakers reported significantly lower relaxation in the GenAI-supported condition as compared to the human-only condition ($p < .05$). Pleasure levels also decreased, though this effect was marginal ( $p = .082$). Conversely, for non-native English speakers, no significant differences in relaxation or pleasure were noted between the GenAI and human-only design sessions. Error bars represent 95% confidence intervals.

*4.2.2 Subgroup: Native Language (native language, non-native language).* For native English speaker participants, our analysis showed that using GenAI for a 25-minute design session decreased their feeling of relaxation. Our analysis found that the mean relaxation score for native English speakers after the designing with GenAI session was 3.375 ($SD = 1.467$), compared with the mean relaxation score of 4.250 ($SD = 1.199$) after the human-only design session. The difference was significant ($Z = -2.481$, $p = .013$). In addition, for native English speakers, there was a marginally significant difference in pleasure between conditions, with pleasure being higher in the human-only condition ($M = 4.833$, $SD = 1.404$) than in the GenAI-supported condition ($mean = 4.208$, $SD = 1.290$, $Z = -1.741, P = 0.082$).

However, for participants who were non-native speakers, there were no significant differences between the two experimental conditions. Condition did not affect other measurements, such as divergent and convergent thinking, self-efficacy, and energy level for either non-native or native speakers.

### 4.3 Qualitative Analysis of Expert Evaluation Comments

[Authors' note: we plan to present design examples in the paper to illustrate the quantitative and qualitative results, but the advertisements contain the client's name which can be associated with the authors' institution. To maintain anonymity, we will not show these design examples in the current manuscript. We will coordinate with the ACs to understand if and how we can show these examples during the later review phase.]

*4.3.1 Perceptions of GenAI-Supported Designs: Increased Creativity and Colorfulness with Potential Drawbacks of Complexity and Confusion.* Experts generally found GenAI-supported designs to be more creative and engaging than human-only designs, frequently praising their "*very creative and appealing*" style (E1). The "*bright colors*" (E3) and "*eye-catching compositions*" (E5) were highlighted as key elements that made these designs stand out. They mentioned these designs potentially could attract viewers in contexts like social media advertisements, as one expert noted, "*I'd guess imagery is much more noticeable as a brief ad*" (E1). The use of vibrant and deep colors was consistently seen as a strength, with descriptions such as "*visually interesting, attractive,*" (E1) and featuring "*flashy colors and cute robots*" (E5) that captured attention. Even when some experts were not enthusiastic about certain designs overall, they still recognized that the colors were "catchy" and visually compelling (E3). Expert evaluators attribute these affects largely to GenAI's innovative use of colors and layouts, which were perceived as more dynamic and engaging compared to the more traditional and conventional approaches found in human-only designs (E1, E3).

However, GenAI-supported designs sometimes led to issues of being too complex and lack of clarity. Experts criticized several GenAI-supported designs for being "*too busy*" (E3, E5) and having "*too many elements*" (E1) that cluttered the visual space, making them difficult to interpret quickly. Comments noted that there were "*too many colors and icons to look at*" (E1), creating a sense of visual overload. Additionally, text legibility was a recurring concern; experts found that the text was often "*not legible,*" which further detracted from the overall effectiveness of these designs (E1).

While GenAI-supported designs mostly followed graphic design norms, they sometimes broke them through unnatural or abnormal compositions. For example, experts were puzzled by choices like images "*overlaying another image that is obstructed*" (E1), frames that appeared "*2D but the actual ad is oriented away from the viewing angle*" (E2), and background elements such as graphs that did not seem to add meaningful content to the poster (E3). Moreover, certain designs were described as "*weird*" or "*uncanny,*" with unusual compositions or elements that felt out of place or unsettling, as one expert said, "*the face looks a bit too uncanny valley for my taste*", and also mentioned another design had "*a weird artifact on the image by [the text] 'innovation'*" (E5).

These aspects suggest that while GenAI-supported designs are perceived as creative and visually striking, they sometimes fail to maintain clarity and are perceived as too busy, potentially limiting their effectiveness in communicating intended messages.

*4.3.2 GenAI-supported Design Broadens Inclusivity, But Sometimes Mis-understanding Social Norms.* Experts noted that GenAI-supported designs effectively showcased diversity, featuring a wide range of technologies and representations of children from various backgrounds. One expert appreciated the "*images showing the diversity of technologies and multiple children*" (E1), and another mentioned, "*representation of diverse teens*" (E3), highlighting the broader inclusivity in these designs. In contrast, some human-only designs relied on more stereotypical imagery, such as the "*overuse of stock imagery of white human hands with white robot hands*" (E3). These observations suggest that GenAI has the potential to expand representation and diversity in these design tasks.

However, experts also pointed out that some GenAI-supported designs misaligned with social norms and could unintentionally convey inappropriate messages. For example, one expert found certain characters problematic, stating that people next to children "*look like adults,*" and if so, "*the hand on the child is immediately disturbing*" (E1). Another comment noted that certain images appeared "*super AI generated, and not in a good way,*" questioning "*why are the teens hugging? And what's in the background?*"' (E3). These critiques highlight that while GenAI designs introduce greater diversity, they can also create uncomfortable or ambiguous imagery, potentially leading to unintended interpretations that deviate from accepted social norms.

## 5 Discussion

> "*The way we define and study creativity has deep implications for how we see ourselves – as more or less agentic beings, as determined by our society and culture or actively shaping it, as different from or similar to the divine*" [50].

GenAI has significantly impacted various aspects of daily life, including enhancing productivity [68], writing [57, 58], interpersonal communication [35], and relationships [34, 76]. In the design industry, GenAI's introduction into creative processes can profoundly affect both designers' working processes and business outcomes, influencing content generation and metrics such as click-through rates [93]. Our study aims to understand GenAI's impact on both design products and designers' inherent creativity. Unlike earlier studies that have focused on using imaginative text-to-text tasks and online evaluations [18], our approach involves a real-world text-to-image task in a controlled lab setting, incorporating expert evaluations from the client's team to assess all eligible designs created by participant designers.

The results indicate that while access to GenAI significantly boosts the perceived creativity and unconventionality ("weirdness") of design products, it does not notably improve their perceived usefulness. In addition, although a brief 25-minute design session with GenAI showed no significant effect on the designers' inherent creativity, mood, or self-efficacy, our subgroup analysis revealed that participants without prior AI design tool experience showed increases in divergent thinking ability, particularly in idea fluency and flexibility. Additionally, native English speakers reported reduced relaxation when using GenAI, indicating that its effects vary based on linguistic background and prior tool exposure.

These findings highlight the complex role of GenAI in shaping design products and influencing designer creativity, providing additional insights as the design industry moves towards more integrated use of GenAI.

### 5.1 Novelty vs. Usefulness: Decoupling GenAI's Impact on Creative Design Outputs

Two commonly used dimensions—novelty and usefulness—have long been used to define creative outputs [61, 62, 74]. Our findings introduce a new perspective on the decoupling of these two dimensions of creativity in the context of GenAI-supported text-to-image tasks. While GenAI-supported designs were perceived as more novel—marked by improved creativity and unconventionality—they were not necessarily considered more useful than those produced under the human-only condition. Although GenAI is able to blend disparate concepts into highly novel ideas, often leading to unconventional and imaginative outputs mentioned in previous literature such as "avocado chairs" and "snail harps" [20, 106], the real-world perceived utility of these creative outputs often remains unknown.

This discrepancy illustrated in our study invites a critical examination of GenAI's role in practical design settings. Although GenAI CSTs are touted for enhancing the design process and improving designers' self-reported performance in research [23], there is a scarcity of external evaluations that affirm the practical utility of the designs in a real-world setting. One prior research using the prediction of a classifier has pointed out that exposure to GenAI's ideas can make people's ideas "*different but not better*" [10]. Our findings join the discussion by challenging the prevailing marketing claims and prior research [43, 84] which suggests that GenAI can enhance the quality of creative outputs. Our expert evaluations show no perceived increase in the utility of a GenAI-supported common design task, a social media advertisement.

The question then arises: does reliance on GenAI foster a misleading expectation of its benefits? To address this, it is crucial to measure not only the outputs of GenAI tools based on designers' self-assessments but also the acceptance and impact of these designs in real-world scenarios. Our planned follow-up study aims to deploy the designs generated in this research on social media platforms, measuring actual user engagement and click-through rates to provide a concrete assessment of their effectiveness and appeal to real audiences.

### 5.2 Supporting Human Creativity in the Age of GenAI

The integration of GenAI into CSTs marks a transformative shift from traditional CSTs, especially with its potential capacity to automate the entire creative process from ideation to prototyping. While previous research has shown the promise of these tools in enhancing individual idea diversity [16, 100], research also found GenAI can homogenize the ideas generated by different users [9]. Furthermore, while automation can boost productivity, it may limit human agency and lead to creative fixation [6, 24] where designers might struggle to venture beyond certain ingrained ideas. Such fixation could reduce the diversity of concepts generated by individuals, and influence inherent human creativity [9].

Our study shows that while GenAI does not consistently affect all users' inherent creativity, it does improve first-time GenAI users' divergent thinking as measured by idea fluency and flexibility, likely because GenAI exposes them to more novel and unconventional ideas. This suggests that GenAI could potentially enhance creativity in users unfamiliar with AI tools, broadening their creative thinking abilities. Conversely, participants with prior GenAI experience did not show significant improvement in divergent thinking after a 25-minute design session, although their designs were rated higher in creativity. This raises questions about whether more creative individuals are drawn to use GenAI or if their creativity is boosted by its use. Future research should be conducted to understand the causal relationship between the two.

Creativity holds intrinsic value, allowing people to express unique human perspectives, emotions, and experiences [72]. Through creative endeavors like writing, art, music, and problem-solving, individuals can explore and articulate who they are [86] alongside providing a sense of human authenticity and fulfillment [50, 91]. As we integrate GenAI into CSTs, it is crucial to discern whether these technologies support or inhibit our creative expressions and our inherent creative abilities. To understand GenAI's long-term effects on creativity, we believe that longitudinal studies are needed to understand whether it truly enhances or inadvertently constrains human creative ability over time.

### 5.3 Supporting Diverse Needs from Different User Groups

Our findings indicate that native speakers experienced diminished relaxation and pleasure when using GenAI to support their design task. We propose that this is likely because of cultural factors. Raised in individualistic Western cultures, native English speakers may display a strong preference for their own ideas, reflecting individualist societies' value on personal autonomy and self-expression [39]. This might lead to a preference for self-generated ideas over those produced by GenAI, partly explained by the "not invented here" syndrome [49], which denotes a reluctance to embrace ideas or products developed outside one's immediate familiar context.

On the other hand, non-native speakers from collectivist cultures might be more open to collaboratively generated ideas [66], including those suggested by AI, due to a cultural emphasis on community and cooperation [19]. This difference in acceptance may explain why native English speakers experienced reduced relaxation and pleasure when using GenAI, potentially perceiving their role more as directors in the AI collaboration rather than partners.

Additionally, our study reveals that GenAI tools enhance divergent thinking among new users, suggesting that initial exposure to AI can stimulate divergent thinking. However, this effect was not observed in individuals with prior experience with such tools, highlighting that the novelty of GenAI may particularly inspire creativity among newcomers.

In light of these insights, there is a significant opportunity for GenAI tool developers to tailor AI tools to meet the diverse needs of different user groups. This could include designing intelligent interfaces that encourage exploration and idea generation for new users while offering more sophisticated controls for experienced users, or developing features that offer collaboration for users from individualist cultures while still providing agency and control. Such an approach would not only improve user experience but also ensure that the tools effectively support creative expression across various cultural and experiential backgrounds.

Future research should explore how cultural and other group factors influence the adoption and use of AI in creative processes, moving beyond simple categorizations of novice versus expert [10, 68]. Exploring user groups from different cultural backgrounds, collaboration patterns, linguistic habits, etc., will deepen our understanding of how CSTs affect creativity and support the human-AI co-creation process. This research is vital for developing GenAI technologies that genuinely enhance human creativity in diverse global contexts.

## 6 Limitation and Future Work

One limitation of our study was the reliance on expert evaluations to assess the quality of design products. While heuristic expert evaluations can serve as a proxy of usefulness [54], they may not fully align with real-world performance metrics, such as click-through rates on social media platforms [29]. To address this, future studies should consider deploying the designs in actual social media campaigns to gather empirical data on user engagement. Additionally, crowdsourcing evaluations could provide a broader perspective on the appeal and effectiveness of these designs, complementing expert opinions with user feedback.

Another constraint was our use of the GPT-4 model powered by DALL-E2 for text-to-image generation, rather than employing CSTs powered by multimodal language models specifically designed to enhance creative thinking. While ChatGPT was chosen for its prevalence and accessibility, recent research has shown that the text-to-image model possesses biases [36, 102]. Future research should investigate the variable impacts of text-to-image GenAI models on creativity and whether design outputs are influenced by different models or not.

## 7 Conclusion

As GenAI becomes more integrated into design workflows, it is crucial to understand how these tools shape both the designers using these tools and the final outputs. Through a within-subjects experiment involving 36 designers, we found that designs created with GenAI support were perceived by expert evaluators as more creative and unconventional. However, these GenAI-supported designs did not show significant improvements in visual appeal, usefulness, or brand alignment compared to designs produced by humans only, suggesting a decoupling of novelty and usefulness—two key components of creativity. Furthermore, while GenAI did not significantly impact the designers' overall creative thinking abilities, our findings highlighted differences among user groups. Designers new to GenAI tools experienced gains in divergent thinking, such as idea fluency and flexibility, while those with prior AI experience did not. Native English speakers reported lower relaxation levels when using GenAI, potentially reflecting cultural factors that influence how different groups interact with GenAI.

These findings suggest that while GenAI can enhance certain aspects of creativity by generating novel and unconventional ideas, it does not automatically translate to more useful outputs. This raises important questions about the role of GenAI in creative industries. We advocate for the development of next-generation GenAI creative support tools that prioritize human creativity and cater to people from different use groups.